\documentclass[11pt]{article}
\usepackage{graphicx,subfigure}
\usepackage{cite}
\usepackage{amsmath,bm}
\usepackage{amsfonts}
\usepackage{parskip}

\usepackage[colorlinks=true,linkcolor=blue,citecolor=blue ,urlcolor=blue]{hyperref}
\usepackage{enumerate}
\usepackage{subfigure}
\usepackage{rotating}
\usepackage{indentfirst}

\title{Entanglement of extremal density matrices of 2-qubit Hamiltonian with Kramers degeneracy}

\author{A. Figueroa, O. Casta\~nos, R. L\'opez-Pe\~na \\
Instituto de Ciencias Nucleares\\
Universidad Nacional Aut\'onoma de M\'exico \\
Apdo. Postal 70-543 M\'exico 04510 D.F.}

\date{}

\usepackage{secdot}
\sectiondot{subsection}
\sectiondot{subsubsection}

\begin{document}

\maketitle

\begin{abstract}

We establish a novel procedure to analyze the entanglement properties of extremal density matrices depending on  the parameters of a finite dimensional Hamiltonian. It was applied to a general 2-qubit Hamiltonian which could exhibit Kramers degeneracy. This is done through the extremal density matrix formalism, which allows to extend the conventional variational principle to mixed states. By applying the positive partial transpose criterion in terms of the Correlation and Schlienz-Mahler matrices on the extremal density matrices, we demonstrate that it is possible to reach both pure and mixed entangled states, changing properly the parameters of the Hamiltonian. For time-reversal invariant Hamiltonians, the extremal pure states can be entangled or not and we prove that they are not time-reversal invariants. For extremal mixed states we have in general 5 possible cases: three of them are entangled and the other two separable. 



\end{abstract}

\maketitle


\clearpage

\section{Introduction}
Recently we have extended the conventional variational method to density matrices of a qudit system. This was done by means of a Lagrange multipliers approach~\cite{figueroa}, using an algebraic procedure~\cite{figueroa2} and through the geometric formulation of quantum mechanics~\cite{figueroa3}. We have shown that, for a finite system the mean value of the Hamiltonian operator $ \hat{H} $ (or any observable) achieves its critical values under the condition $[ \hat{\rho} \,,\, \hat{H} ] = \boldsymbol{0}$, which it is equivalent to the stationary solution of the von Neumann equation. We call extremal density matrices to the states that fulfill this condition. They provide an extremal description of the mean values of the Hamiltonian, and in the case of restricting them to pure states, one recovers the energy spectrum. So, besides of being an alternative tool to find either the eigensystem or detect level crossings in the Hamiltonian without computing its eigenvalues, one obtains information of mixed states which minimize the mean value of the energy. 


Among other methods to detect entanglement in bipartite systems~\cite{guhne}, it has been shown that if the entire system is separable, the partial transposition operation has the property of preserving the positive definiteness of the density matrix. Hence, the necessary condition for separability of a finite dimensional state is the positivity of its partial transposition, also commonly referred as PPT criterion~\cite{horodecki}. For the 2-qubit state and the tensorial product of a qubit times a qutrit, the PPT criterion is also a sufficient one~\cite{peres}. 

The present article arises from the concern about the entanglement of extremal density matrices, associated to time-reversal invariant Hamiltonian matrices denoting 2-qubit general systems. A Hamiltonian with this type of time-reversal symmetry can be constructed following the procedure indicated by Haake~\cite{haake}. As we shall shown, the main novelty is that, by applying the extremal density matrix procedure together with the PPT criterion, one can obtain mixed separable or mixed entangled states by changing properly the parameters of the time-reversal invariant Hamiltonian. Extending this result further, at least for $4$ and $6$ dimensional Hilbert spaces, it will be possible to obtain mixed separable and mixed entangled states tuning the parameters of the Hamiltonian, whether it has a specific symmetry or not.

The paper is organized as follows. We start with a summary of the procedure to determine extremal density matrices of a finite dimensional Hamiltonian. In section \ref{kramers}, the Kramers degeneracy or time-reversal invariance is reviewed and the general form of a $ 4\times 4 $ Hamiltonian with this symmetry is given. The PPT criterion is established in section \ref{pptsec}, and its connection with the semi-positivity conditions on the extremal density matrices is shown. This connection is given by the Correlation and Schlienz-Mahler matrices~\cite{schlienz,mahler}. Finally, in section \ref{examples}, our method is applied to the general four-dimensional Hamiltonian which exhibits Kramers degeneracy.

\section{Extremal density matrices}\label{count}

The space of Hermitian matrices can be stratified by means of the Partial Flag Manifolds $ F(d;\, m_1,m_2,\ldots,m_k) $ and the number of different strata is equal to the partition function $ p(d) $ (see \ref{diophsol}). If we denote by $r$ the dimension of $ F(d;\, m_1,m_2,\ldots,m_k) $, then any quotient of unitary groups $ \hat{U} \in F(d;\, m_1,m_2,\ldots,m_k) $, is specified by $ r $ real parameters (see Table~\ref{tabdim}). Consequently, the dimension of $ \hat{H} $, i.e., the number of independent real parameters needed to specify $ \hat{H} $, is given by the formula~\cite{keller}
\begin{eqnarray}
{\rm dim} (\hat{H}) = d^2 - \biggl( \sum_{j=1}^{k} m_j^2 - k \biggr) \, , \label{dim}
\end{eqnarray}
where $ k $ is the number of different eigenvalues of $ \hat{H} $ and $ \{ m_j \} $ denote their algebraic multiplicities, such that $ m_1+m_2+ \cdots + m_k = d $. The quantity in parentheses is the codimension of $ \hat{H} $, $ {\rm codim} (\hat{H}) $, and represents the number of conditions to be fulfilled for a level crossing~\cite{caspers}. 
\begin{table*}[h!]
\centering
\caption{ \label{tabdim}
Dimension, Codimensions and Partial Flag Manifolds $ F(d;\, m_1,m_2,\ldots,m_k) $ for the unitary orbits of $ \hat{H} $. It is supposed that $ \alpha > \beta > \gamma > \delta $. If they are given explicitly, the dimension $ {\rm dim} (\hat{H}) $ is decreased by $k$ and it is denoted by $r$, which is also the dimension of the Manifold. }
\medskip
\resizebox{0.99  \textwidth}{!} {
\begin{tabular}{ | c | c c c c c c |}
\hline
\hline
\textbf{Hilbert space} & \textbf{Hermitian} & \textbf{Eigenvalue} & \textbf{Codimension} & \textbf{Dimension}  & \textbf{Partial Flag} & \textbf{Manifold} \\
\textbf{dimension} & \textbf{diagonal} & \textbf{parameters} &  $ {\rm codim} (\hat{H}) $ &  $ {\rm dim} (\hat{H}) $  & \textbf{Manifold} & \textbf{ dimension } \\
$ \boldsymbol{ d} $ & \textbf{representation} & k & $\sum_{j=1}^{k} m_j^2 - k$  & $ d^2 - {\rm codim} (\hat{H}) $ & $ F(d;\, m_1,m_2,\ldots,m_k) $ & $ r = {\rm dim} (\hat{H}) - k $ \\
\hline
\hline
\hline
    	    & ${\rm diag} (\alpha , \alpha )$ & 1 & 3 & \textbf{1} & Point & \textbf{0}  \\
            \textbf{2} & $ {\rm diag} (\alpha , \beta )$ & 2 & 0 & \textbf{4} & $ U(2)/[U(1) \times U(1) ]  $ & \textbf{2} \\
\hline
    	    & ${\rm diag} (\alpha , \alpha , \alpha )$ & 1 & 8 & \textbf{1} & Point & \textbf{0} \\
\textbf{3} & ${\rm diag} (\alpha , \beta , \beta )$ & 2 & 3 & \textbf{6} & $ U(3)/[U(1) \times U(2) ]  $  & \textbf{4} \\
    		& $ {\rm diag} (\alpha , \beta , \gamma )$ & 3 & 0 & \textbf{9} & $ U(3)/[U(1) \times U(1) \times U(1) ]   $ & \textbf{6} \\
\hline
    		& ${\rm diag} (\alpha , \alpha , \alpha, \alpha )$ & 1 & 15 & \textbf{1} & Point & \textbf{0} \\
   		& $ {\rm diag} (\alpha , \beta , \beta, \beta )$ & 2 & 8 & \textbf{8} & $ U(4)/[U(1) \times U(3) ]  $ & \textbf{6} \\
\textbf{4}  & $ {\rm diag} (\alpha , \alpha , \beta, \beta )$ & 2 & 6 & \textbf{10} & $ U(4)/[ U(2) \times |U(2) ]  $ & \textbf{8} \\
    		& $ {\rm diag} (\alpha , \beta , \gamma , \gamma )$ & 3 & 3 & \textbf{13} & $ U(4)/[U(1) \times U(1) \times U(2) ]  $ & \textbf{10} \\
    		& $ {\rm diag} (\alpha , \beta , \gamma , \delta )$ & 4 & 0 & \textbf{16} & $ U(4)/[U(1) \times U(1) \times U(1) \times U(1) ]  $ & \textbf{12} \\
\hline
\end{tabular}}
\end{table*}

The Rayleigh quotient $e_H(\psi)$ of a Hermitian matrix $\hat{H}$ is 
\begin{eqnarray}
e_H(\psi) := \frac{\langle \psi|  \hat{H}  |\psi \rangle}{\langle \psi|  \psi \rangle} \, ,  \nonumber
\end{eqnarray}
where $|\psi \rangle$ is a $d$-dimensional complex vector. The numerical range $ W(\hat{H}) $ is the set of all possible Rayleigh quotients $ e_H(\psi)$ over the unit vectors. It is a closed interval on the real axis and the eigensystem of $\hat{H}$ is associated to the critical points of $ e_H(\psi)$. Thus $W(\hat{H})$ is the convex hull of the eigenvalues~\cite{horn}.

In the density matrix formalism, the numerical range of the Hamiltonian (or any Hermitian operator) can be identified with $ \langle \hat{H} \rangle = {\rm Tr}( \hat{H}\, \hat{\rho} ) $ and, by restricting to variations along the unitary orbits, it can be shown that under the condition $[ \hat{\rho} \,,\, \hat{H} ] = \boldsymbol{0} $ the Rayleigh quotient achieves its critical values at $  \langle \hat{H} \rangle^c = {\rm Tr} ( \hat{H} \,  \hat{\rho}^c ) $, where $ \hat{\rho}^c $ denotes an extremal density matrix commuting with $ \hat{H} $~\cite{figueroa2,figueroa3}. 

It is known that two Hermitian operators with vanishing commutator share a common eigenbasis. The difference between degenerate and non-degenerate cases of $ \hat{H} $ is that, in the former case, not all its eigenvectors are necessarily eigenvectors of $ \hat{\rho}^c $. However, it is generally expected that for a given Hamiltonian $ \hat{H} $, the dimension of an unknown $ \hat{\rho}^c $ can be determined up to the same Partial Flag Manifold dimension of $ \hat{H} $, i.e., from Table~\ref{tabdim}, the number of parameters describing $ \hat{\rho}^c $ is
\begin{eqnarray}
{\rm dim} (\hat{\rho}^c) = r \, . \label{dimr}
\end{eqnarray}
Then the number of free parameters $n$ of $ \hat{\rho}^c $ is given by
\begin{eqnarray}
n = (d^2-1) - r \, . \label{dimn}
\end{eqnarray}

An algebraic proof of these results is found by using $[ \hat{H}, \hat{\rho} ]=0$, the properties of the generators of the $su(d)$ algebra, the tangent vectors of $ \hat{H} $ at the identity and the Gram matrix $ \boldsymbol{G} $ formed with them~\cite{figueroa2}. Therefore, $r = {\rm rank} \, \boldsymbol{G} $ also determines the dimension of the tangent space of the Manifolds (see Table \ref{tabdim}) and is an alternative tool to detect level crossings without computing the eigenvalues of $\hat{H} $  .

\section{Hamiltonians with Time-Reversal Invariance}\label{kramers}

Unitary transformations leaving invariant a Hamiltonian matrix give rise conserved quantities. Anti-unitary transformations sometimes increase the degree of degeneracy as it is the case for the time-reversal invariance, described by an operator $ \hat{T} $ which satisfies $ \hat{T}^2 = - \hat{I} $ and $ [ \hat{T} ,\hat{H} ] = \boldsymbol{0} $. For half-integer total spin $S=(2N-1)/2$ ($ N \in \mathbb{N} $, even-dimensional Hilbert space $d=2N$), time-reversal invariance implies Kramers degeneracy, i.e., pairs of energy levels of $ \hat{H} $ are degenerated~\cite{haake}. For any state vectors $ \{ | \phi \rangle , \, | \psi \rangle \} $ and complex numbers $ \{ c_1 ,\, c_2 \} $, $ \hat{T} $ has the properties~\cite{jordan}:
\begin{enumerate}[i)]
\item (fermionic condition) $ \hat{T}^2 = - \hat{I} $,
\item (anti-linearity) $ \hat{T} ( \, c_1 | \psi \rangle  + c_2 | \phi \rangle \, ) = c_1^{\ast} \, \hat{T} \, | \psi \rangle  +  c_2^{\ast} \, \hat{T} \, | \phi \rangle$,
\item (anti-unitarity) $ \langle \hat{T} \psi | \hat{T} \phi \rangle = \langle \psi | \phi \rangle^{\ast} $. 
\end{enumerate}

From these properties, it follows that $ \hat{T} $ has an inverse $ \hat{T}^{-1} $, preserves the norm, $| \hat{T} \psi \rangle$ is unique and orthogonal to $| \psi \rangle$, and consequently, $\hat{T}$ has no eigenvectors. Three implications can be deduced~\cite{rosch}: All eigenvalues of $\hat{H}$ are doubly degenerate (Kramers degeneracy); the corresponding Hilbert space $ \mathbb{H}_d $ can not be decomposed into invariant subspaces with respect to $ \hat{T} $; and a symmetry-adapted orthonormal basis for $ \mathbb{H}_d $ exists and it takes the form $\{ | \psi_1 \rangle , | \hat{T} \psi_1 \rangle, | \psi_2 \rangle, | \hat{T} \psi_2 \rangle , \ldots, | \psi_N \rangle , | \hat{T} \psi_N \rangle\} $. 

In this basis, for $N=2$ ($d=4$) case, the general form of the traceless Hamiltonian~\cite{haake},
\begin{eqnarray}
\hat{H} & = & \left(
\begin{array}{cccc}
 \beta  & 0 & \gamma -i s & -\epsilon -i \delta  \\
 0 & \beta  & \epsilon -i \delta  & \gamma +i \sigma  \\
 \gamma + i s  & \epsilon +  i \delta & -\beta  & 0 \\
-\epsilon +  i \delta  & \gamma -i \sigma  & 0 & -\beta 
\end{array}
\right) \, , \label{ham}
\end{eqnarray}
presents Kramers degeneracy if $ s = \sigma $ and consequently, it is double degenerate. On the other hand, we consider broken the time-reversal invariance  by taking $ s = - \sigma $, thus $ \hat{H} $ is non degenerate. 

\section{Separability criterion for two mixed qubits}\label{pptsec}

A basis for the $2$-dimensional matrices is given by the identity and Pauli matrices,
\begin{eqnarray}
\hat{\sigma}_0 = \left(
\begin{array}{cc}
1 & 0 \\
0 & 1
\end{array} \right) , \, 
\hat{\sigma}_1 = \left(
\begin{array}{cc}
0 & 1 \\
1 & 0
\end{array} \right) , \, 
\hat{\sigma}_2 = \left(
\begin{array}{cc}
0 & - i \\
i & 0
\end{array} \right) , \, 
\hat{\sigma}_3 = \left(
\begin{array}{cc}
1 & 0 \\
0 & -1
\end{array} \right) \, , \nonumber
\end{eqnarray}
with the product rule $ \hat{\sigma}_{i} \, \hat{\sigma}_{j} = i \, \epsilon_{ijk} \, \hat{\sigma}_{k} + \delta_{ij} \hat{\sigma}_0 $, and orthogonality relation $ {\rm Tr} ( \hat{\sigma}_{j} \, \hat{\sigma}_{k} ) = 2 \delta_{jk} $. An arbitrary Hamiltonian $\hat{H} $ and single-qubit state $ \hat{\rho} $ can be represented as
\begin{eqnarray}
\hat{H} & = &  \frac{1}{2} \sum_{k=0}^{3} h_{k} \, \hat{\sigma}_{k}   \, , \\
\hat{\rho} & = &  \frac{1}{2}  \sum_{k=0}^{3} r_{k} \, \hat{\sigma}_{k}  \, ,
\end{eqnarray}
with the identifications $ h_{k} = {\rm Tr} ( \hat{H} \, \hat{\sigma}_{k} ) , \,  r_{k} = {\rm Tr} ( \hat{\rho} \, \hat{\sigma}_{k} ) $ and $ r_{0} = {\rm Tr} ( \hat{\rho} ) = 1 $.

Similarly, in the 2-qubit case $\hat{H} $ and $\hat{\rho} $ can be parametrized as 
\begin{eqnarray}
\hat{H} = \frac{1}{4}  \sum_{p=0}^{3} \sum_{q=0}^{3} h_{p\,q} \, \hat{D}_{p,q}  \, , \label{hamiltonian} \\
\hat{\rho} = \frac{1}{4}  \sum_{p=0}^{3} \sum_{q=0}^{3} r_{p\,q} \, \hat{D}_{p,q} \label{denmatgen} \, ,
\end{eqnarray}
where $ \hat{D}_{p,q} = \hat{\sigma}_p \otimes \hat{\sigma}_q $, $ h_{p\,q} = {\rm Tr} ( \hat{H} \, \hat{D}_{p,q} ) $, $ r_{p\,q} = {\rm Tr} ( \hat{\rho} \,\hat{D}_{p,q} ) $ and $ r_{0 0} = {\rm Tr} ( \hat{\rho} ) = 1 $. In matrix form, $ \hat{\rho} $ can be written in terms of $ 2 \times 2 $ matrices,
\begin{eqnarray}
\hat{\rho} &=& 
\left(
\begin{array}{cc}
F & G \\
P & Q
\end{array}
\right)  , \, \label{rhogen}  
\end{eqnarray}
where $ P = G^{\dagger} $ and
\begin{eqnarray}
F &=&
\frac{1}{4} \left(
\begin{array}{cc}
 r_{03}+r_{30}+r_{33}+1 & r_{01}+r_{31}-i(r_{02}+ r_{32}) \\
 r_{01}+r_{31}+i(r_{02}+ r_{32})  & -r_{03}+r_{30}-r_{33}+1
\end{array}
\right) ,  \, \nonumber \\
G &=&
\frac{1}{4} \left(
\begin{array}{cc}
r_{10}+r_{13}-i(r_{20}+ r_{23}) & r_{11}-r_{22} -i (r_{12} + r_{21}) \\
r_{11}+r_{22} +i( r_{12}-r_{21}) & r_{10}-r_{13}- i( r_{20}- r_{23}) 
\end{array}
\right) , \, \nonumber \\
Q &=&
\frac{1}{4} \left(
\begin{array}{cc}
r_{03}-r_{30}-r_{33}+1 & r_{01} - r_{31}- i( r_{02}- r_{32}) \\
r_{01} - r_{31} + i( r_{02}- r_{32}) & -r_{03}-r_{30}+r_{33}+1
\end{array}
\right) \, . \nonumber
\end{eqnarray} 

The basis matrices $ \{ \hat{D}_{p,q} = \hat{\sigma}_p \otimes \hat{\sigma}_q \} $  fulfill the orthogonality condition
\begin{eqnarray}
{\rm Tr} ( \hat{D}_{j,k} \, \hat{D}_{m,n} ) = 4 \delta_{jm} \, \delta_{kn} \, ,
\end{eqnarray}
and they comply with the multiplication rules
\begin{eqnarray}
\hat{D}_{p,0} \, \hat{D}_{0,q} & = & \hat{D}_{0,q} \, \hat{D}_{p,0} = \hat{D}_{p,q}  \, , \nonumber \\
 \hat{D}_{i,0} \, \hat{D}_{j,0} & = & i \, \epsilon_{ijk} \, \hat{D}_{k,0} + \delta_{ij} \hat{D}_{0,0} \, , \\
 \hat{D}_{0,i} \, \hat{D}_{0,j} & = & i \, \epsilon_{ijk} \, \hat{D}_{0,k} + \delta_{ij} \hat{D}_{0,0} \, . \nonumber
\end{eqnarray}
By means of the previous expressions, the following commutators can be calculated: 
\begin{eqnarray}
\lbrack \hat{D}_{j,0} \, , \,  \hat{D}_{m,n} \rbrack & = & 2 i \, \epsilon_{jnq} \, \hat{D}_{m,q} \, , \nonumber \\
\lbrack \hat{D}_{0,j} \, , \,  \hat{D}_{m,n} \rbrack & = & 2 i \, \epsilon_{jmq} \, \hat{D}_{q,n} \, , \\
\lbrack \hat{D}_{i,j} \, , \,  \hat{D}_{p,q} \rbrack & = & 2 i \, \biggl( \epsilon_{jql} \, \delta_{ip} \hat{D}_{l,0} +  \epsilon_{ipk} \, \delta_{qj} \hat{D}_{0,k}  \biggr) \, . \nonumber
\end{eqnarray}

The parameterisation~(\ref{denmatgen}) is an abbreviation of the Fano form~\cite{bengtsson,schlienz},
\begin{eqnarray}
\hat{\rho}  &=& \frac{1}{4} \left( \hat{I}_4 + \sum_{p=1}^{3} r_{p\, 0} \, \hat{D}_{p,0} +  \sum_{q=1}^{3} r_{0 \,q} \,  \hat{D}_{0,q} + \sum_{p=1}^{3} \sum_{q=1}^{3} C_{p\,q} \,  \hat{D}_{p,q}  \right) \, , \label{fanoform}
\end{eqnarray}
where the two Bloch vectors 
\begin{eqnarray}
\tau^{A} & = & ( r_{10},r_{20},r_{30} ) , \quad
\tau^{B} = ( r_{01},r_{02},r_{03} ) \, ,
\end{eqnarray}
determine the properties of the individual qubits $A$ and $B$. For $s, t = 1,2,3 $, the matrix $ C_{s\,t} = r_{s\,t}$ is given by
\begin{eqnarray}
\boldsymbol{C} = \left(
\begin{array}{cccc}
r_{11} & r_{12} & r_{13} \\
r_{21} & r_{22} & r_{23} \\
r_{31} & r_{32} & r_{33} 
\end{array}
\right) \, ,
\end{eqnarray}
from which it is possible to build $ M_{s\,t} = C_{s\,t} - r_{s\,0} \,  r_{0 \,t} $,
\begin{eqnarray}
\boldsymbol{M} = \left(
\begin{array}{cccc}
C_{11}-r_{10} \, r_{01} & C_{12}-r_{10} \, r_{02} & C_{13}-r_{10} \, r_{03} \\
C_{21}-r_{20} \, r_{01} & C_{22}-r_{20} \, r_{02} & C_{23}-r_{20} \, r_{03} \\
C_{31}-r_{30} \, r_{01} & C_{32}-r_{30} \, r_{02} & C_{33}-r_{30} \, r_{03} 
\end{array}
\right) \, .
\end{eqnarray}
They are known as the Correlation matrix $ \boldsymbol{C} $ and the Schlienz-Mahler matrix $ \boldsymbol{M} $, respectively~\cite{schlienz}, and together they describe the correlations between both subsystems. If $ \boldsymbol{C} = \boldsymbol{0} $ or $ \boldsymbol{M} = \boldsymbol{0} $ then the state is separable, while for pure states a good measure of entanglement is 
\begin{eqnarray}
\beta = \frac{4}{15}  \, {\rm Tr} ( \boldsymbol{M} ^T \boldsymbol{M}  ) \, , \label{betapar}
\end{eqnarray}
where $ 0 \leq \beta \leq 1$ and, from here on, the upper index $T$ denotes transposition of matrices. The value $ \beta = 1 $ corresponds to the maximal entangled pure state with $\tau^{A} = \tau^{B} = \boldsymbol{0}$, whereas $ \beta = 0$ to a separable one~\cite{mahler}. Also, $ \beta  $ can be interpreted geometrically as a distance between entangled and separable states~\cite{aniello}. Additionally, it was noted in~\cite{quesne,gerdt2} that $ \det\boldsymbol{C} $ and $ \det \boldsymbol{M} $ can be written in terms of $ SU(2)\times SU(2) $ polynomial invariants of third and fourth degrees, respectively.

In order to extend the applications of the extremal mixed density matrices, we consider the PPT criterion in the four dimensional case given in terms of the positivity conditions of the density matrix together with the Correlation and Schlienz-Mahler matrices. 

Taking into account~(\ref{rhogen}), the partial transpose of $ \hat{\rho} $ with respect to the A subsystem, denoted as $ \hat{\rho}^{PT_A} $, and similarly for B, are given by~\cite{bengtsson}
\begin{eqnarray}
\hat{\rho}^{PT_A} &=& 
\left(
\begin{array}{cc}
F & P \\
G & Q
\end{array}
\right)  , \quad
\hat{\rho}^{PT_B} = 
\left(
\begin{array}{cc}
F^T & G^T \\
P^T & Q^T
\end{array}
\right) \, .
\end{eqnarray} 

The PPT criterion is a necessary condition, for the joint density matrix of the 2-qubits $A$ and $ B $ subsystems to be separable. In the $ 4 $ and $ 6 $ dimensional cases the condition is also sufficient~\cite{horodecki}, i.e., if the partial transposed density matrix $\hat{\rho}^{PT_A} $ ($\hat{\rho}^{PT_B} $) is positive definite, the state $ \hat{\rho}$ is separable, or else, the state $ \hat{\rho}$ is entangled if its partial transposition is not positive definite. Notice that $\hat{\rho}^{PT_A} $ and $\hat{\rho}^{PT_B} $ share the same characteristic polynomial and then the indices A and B can be omitted. This entails that, in the 2-qubit case, the PPT criterion and the $ \boldsymbol{C} $ and $ \boldsymbol{M} $ matrices are related through the coefficients $ \{ a_2^{PT} ,\, a_3^{PT} ,\, a_4^{PT} \} $ of the characteristic polynomial of the $ 4 \times 4 $ partial transposed density matrix $ \hat{\rho}^{PT}$ from $ \hat{\rho}$ as~\cite{gerdt2}
\begin{eqnarray}
0 \leq \enspace a_2^{PT} & = & a_2 \enspace \leq \frac{3}{8}  \, , \nonumber \\
0 \leq \enspace a_3^{PT} & = & a_3 + \frac{1}{4} \det \boldsymbol{C} \enspace \leq \frac{1}{16} \, , \label{pospt} \\
0 \leq \enspace a_4^{PT} & = & a_4 + \frac{1}{16} \det \boldsymbol{M} \enspace \leq \frac{1}{256} \, , \nonumber
\end{eqnarray}
such that any separable state represented by $ \hat{\rho} $ must fulfill those inequalities. Conversely, if an inequality is violated then the state $ \hat{\rho}$ is entangled. Hence, given a $ 4\times 4 $ extremal density matrix, one can know immediately whether the states that they represent are separable or not.

\section{Entaglement for a 2-qubit extremal density matrices }\label{examples}
In order to apply the discussed separability criteria together with the extremal density matrix approach, in this section we will consider the general parametrization (\ref{fanoform}) and two cases in the Hamiltonian~(\ref{ham}): The non-degenerate case when $ s = - \sigma $, and the Kramers degeneracy case when $ s = \sigma $. 

\subsection{Non-degenerate case $ ( s = - \sigma ) $}

In this case, the matrix Hamiltonian does not commute with the time-reversal antiunitary operator $\hat{T}$. Under the condition $[ \hat{\rho} \,,\, \hat{H} ] = \boldsymbol{0} $, the density matrix commuting with $\hat{H}$ takes the following values of $r_{p q}$
\begin{eqnarray}
r_{20} &=& \frac{r_{11} \left(\delta ^2+\epsilon ^2\right)-\delta \, \sigma \, r_{10} }{\gamma  \delta } , \, r_{30}=\frac{\beta \, r_{10}}{\gamma } , \, r_{01}= \frac{\delta \, r_{02}}{\epsilon } , \, r_{12} = \frac{\epsilon \,  r_{11} }{\delta } , \, r_{21} = \frac{\delta \, r_{10}- \sigma \, r_{11} }{\gamma }  , \nonumber \\ 
r_{22} &=& \frac{\epsilon \, (\delta \, r_{10}- \sigma \, r_{11} )}{\gamma  \delta } , \, r_{31}= \frac{\beta \, r_{11}}{\gamma } , \, r_{32}=  \frac{\beta \, \epsilon \, r_{11} }{\gamma  \delta }, \, r_{03}=  r_{13}=  r_{23}=  r_{33}=0 \, ,
\end{eqnarray}
with $3$ free variables $ r_{10} , r_{02} , r_{11} $. Hence, from Table~\ref{tabdim} and Eq.~(\ref{dimn}), $ \hat{H} $ is non degenerate. 

\begin{figure}
\begin{center}
\includegraphics[width=0.6 \textwidth]{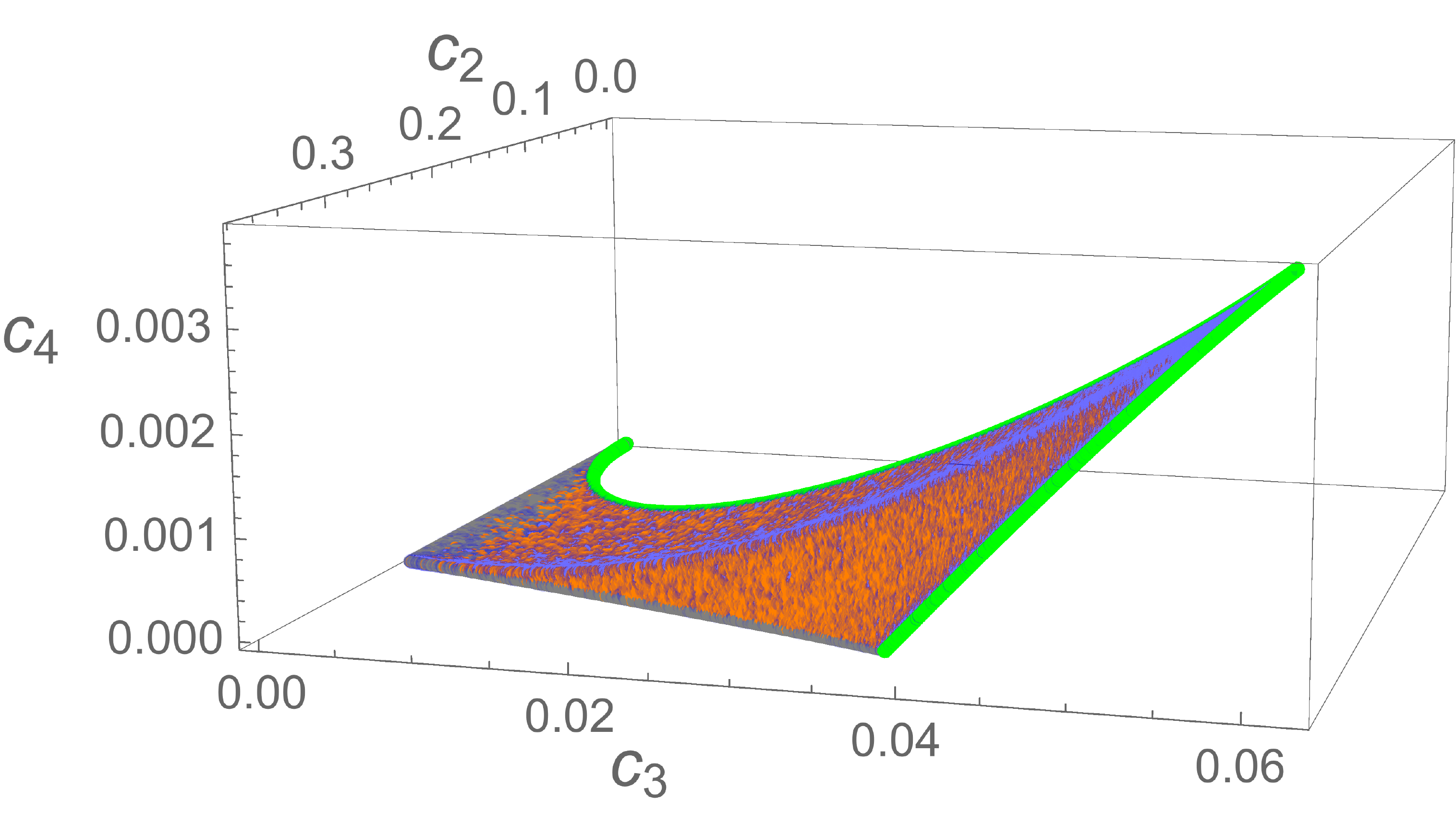}
\end{center}
\caption{ Solid figure which represents the region of $c_2$, $c_3$, and $c_4$ where the positivity conditions of density matrix are satisfied (see \ref{B}). The pure case is associated to $(c_2,c_3,c_4)=(0,0,0)$ while the maximal mixed state corresponds to $(c_2,c_3,c_4)=(3/8,1/16,1/256)$.} 
\label{region}
\end{figure}

These free variables are determined by establishing the system of polynomial equations~(\ref{s2:eq3}), where the constants $ c_2 $, $ c_3 $ and $ c_4 $ must lie inside the allowed region exhibited in Fig.~\ref{region}.  

For the pure case, associated to $c_2 = c_3 = c_4 = 0$, the set of solutions for this polynomial systems are denoted as $ \{ \hat{\rho}^{c}_{1 \pm} ,\, \hat{\rho}^{c}_{2 \pm} \} $, whose respective Bloch vectors, Correlation and Schlienz-Mahler matrices are 
\begin{eqnarray}
\tau^A_{1 \pm} & = & \frac{-1}{\sqrt{\delta ^2+\epsilon ^2}} \biggl( \delta , \,  \epsilon  , \,  0 \biggr) , \quad
\tau^B_{1 \pm} = \pm \frac{1}{E_+} \biggl(  \gamma , \, - (\sqrt{\delta ^2+\epsilon ^2}+\sigma ) , \,  \beta  \biggr) \, , \nonumber \\
\boldsymbol{C}_{1 \pm} & = & \pm \frac{1}{E_+ \, \sqrt{\delta ^2+\epsilon ^2}} \left(
\begin{array}{cccc}
- \gamma  \delta &  - \gamma  \epsilon &  0  \\
 \delta  \left(\sqrt{\delta ^2+\epsilon ^2}+\sigma \right) &  \epsilon  \left(\sqrt{\delta ^2+\epsilon ^2}+\sigma \right) & 0 \\
- \beta  \delta &  - \beta  \epsilon  & 0
\end{array}
\right) \, , \nonumber \\
\tau^A_{2 \pm} & = &  \frac{1}{\sqrt{\delta ^2+\epsilon ^2}} \biggl( \delta , \,  \epsilon  , \,  0 \biggr) , \quad
\tau^B_{2 \pm} = \pm \frac{1}{E_-} \biggl(  \gamma , \,( \sqrt{\delta ^2+\epsilon ^2}- \sigma ) , \, \beta  \biggr) \, , \label{solsnodeg}  \\
\boldsymbol{C}_{2 \pm} & = & \pm \frac{1}{E_- \, \sqrt{\delta ^2+\epsilon ^2}} \left(
\begin{array}{cccc}
 \gamma  \delta &   \gamma  \epsilon &  0  \\
 \delta  \left(\sqrt{\delta ^2+\epsilon ^2}- \sigma \right) &  \epsilon  \left(\sqrt{\delta ^2+\epsilon ^2}-\sigma \right) & 0 \\
 \beta  \delta &   \beta  \epsilon  & 0
\end{array}
\right) \, , \nonumber \\
\boldsymbol{M}_{1 \pm} & = & \boldsymbol{M}_{2 \pm} = \boldsymbol{0} \, , \nonumber 
\end{eqnarray}
where 
\begin{eqnarray}
E_{\pm} & = & \sqrt{\beta ^2+\gamma ^2 +\delta ^2+\sigma ^2+\epsilon ^2 \, \pm \, 2 \sigma  \sqrt{\delta ^2+\epsilon ^2}} \, . \nonumber
\end{eqnarray}

The respective expectation values of the Hamiltonian are
\begin{eqnarray}
{\rm Tr} ( \hat{\rho}^c_{1 \pm } \,\hat{H} ) = \pm E_+ \, , \quad
{\rm Tr} ( \hat{\rho}^c_{2 \pm } \,\hat{H} ) = \pm E_-  \, , \nonumber
\end{eqnarray}
and it is possible to corroborate that $ \{ \hat{\rho}^c_{1\pm}, \, \hat{\rho}^c_{2\pm} \} $ form a complete set of orthogonal states. They are rank one projectors and $ \hat{\rho}^c_{1+} + \hat{\rho}^c_{1-} + \hat{\rho}^c_{2+} + \hat{\rho}^c_{2-} = \hat{I}_{4}$. All of them are separable because $ \boldsymbol{M}_{1 \pm} = \boldsymbol{M}_{2 \pm} = \boldsymbol{0} $. Due to the convex property of the separable states~\cite{bertlmann2} and the unitarily evolution, it follows that $ \hat{H} $ does not generate entangled mixed states.

\begin{figure}
	\centering
	(a) \includegraphics[width=0.4\textwidth]{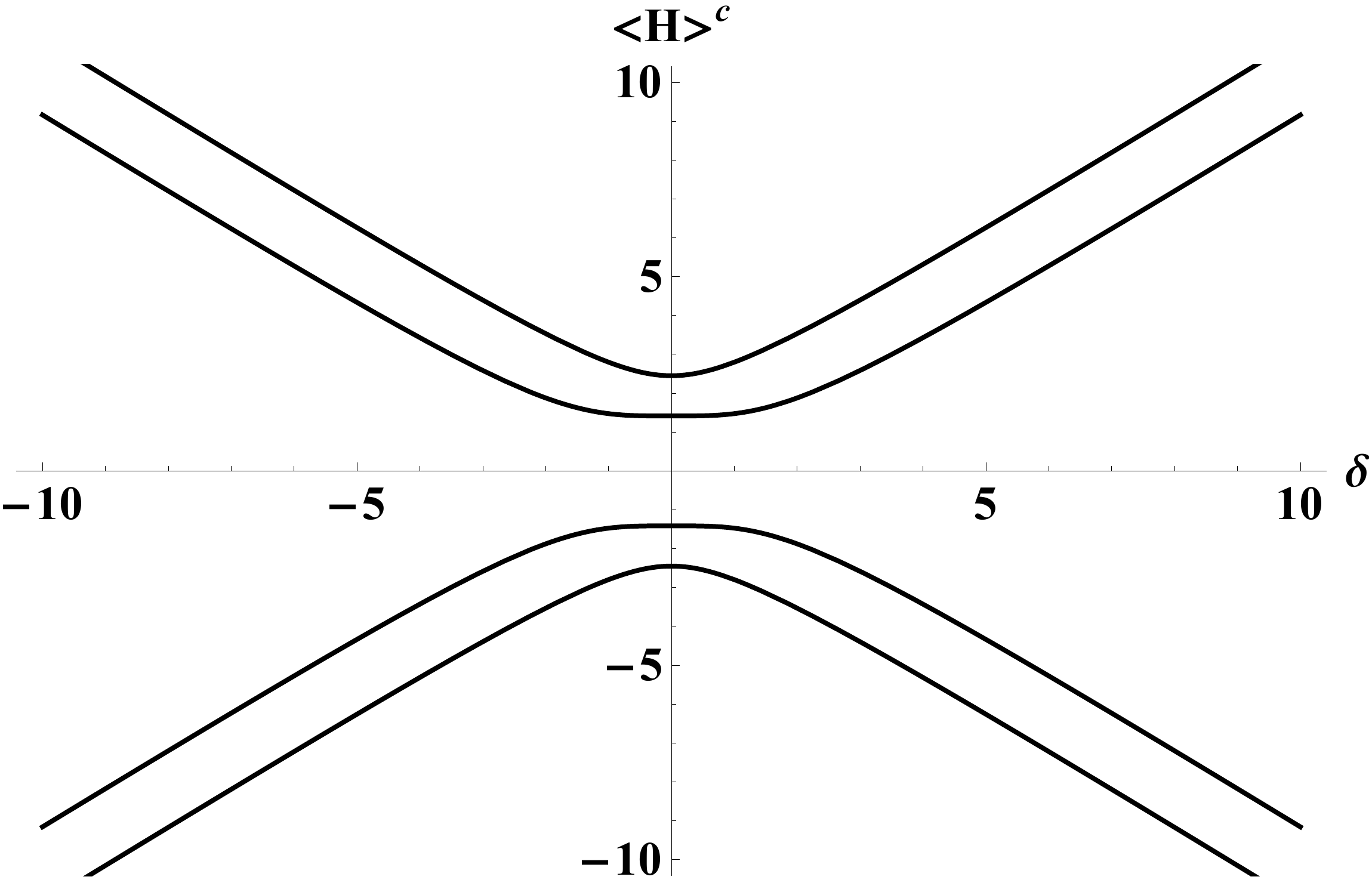}\qquad
	(b) \includegraphics[width=0.4\textwidth]{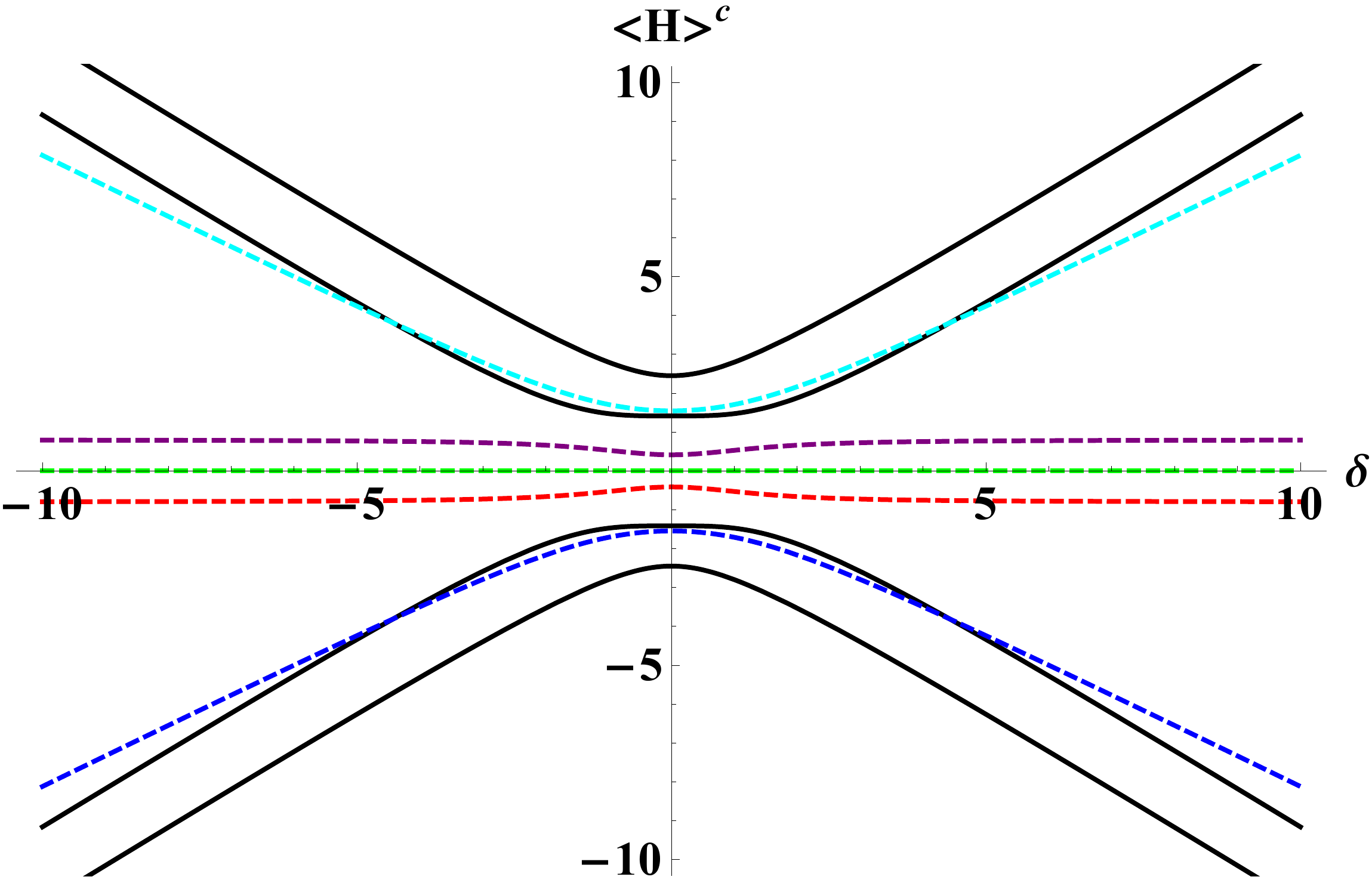}
\caption{ $ \langle \hat{H} \rangle^c $ as a function of $ \delta $ with $\beta = \gamma = \sigma = \epsilon = 1 $. (a) Pure case with $c_2 = c_3 = c_4 = 0$; and (b) Mixed case with $c_2 = 59/200, \, c_3 = 9/400 ,\, c_4 = 81/160 000 $. Black continuous lines represent mean values of $ \hat{H} $ in the pure case. The dotted ones correspond to the mixed case.}
	\label{mixto}
\end{figure}

In order to study the behavior of the expectation values of the Hamiltonian, we consider the parameters $\beta = \gamma = \sigma = \epsilon = 1 $ for two cases: (i) the pure case when one has $c_2 = c_3 = c_4 = 0$, which has four independent solutions for the variables $ r_{10}, r_{20} , r_{11} $. The energy spectra is a function of the parameter $\delta$ and its energy levels are plotted in Fig~\ref{mixto}(a). (ii) The mixed case is established by taking from the region exhibited in Fig.~\ref{region}(c) the values $c_2 = 59/200, \, c_3 = 9/400 ,\, c_4 = 81/160 000 $. One has $6$ extremal expectation values of the Hamiltonian, two of them correspond to $ \langle \hat{H} \rangle^c = 0 $. The results are shown in Fig.~\ref{mixto}(b) with dotted lines. Notice that the extremal expectation values are contained within the minimum and maximum eigenvalues of the Hamiltonian, as it should be.

\subsection{Degenerate case $ ( s = \sigma ) $}

Under the condition $[ \hat{\rho} \,,\, \hat{H} ] = \boldsymbol{0} $, the density matrix commuting with $\hat{H}$ takes the following values for the parameters $r_{p q}$:
\begin{eqnarray}
r_{31} &=& \frac{- \sigma \, r_{02} + \epsilon \, r_{03} + \beta \,  r_{11}}{\gamma },\, r_{32} = \frac{\sigma \, r_{01} - \delta \, r_{03}+\beta \, r_{12}}{\gamma },\, r_{33} = \frac{- \epsilon \, r_{01} + \delta \, r_{02}+\beta \, r_{13}}{\gamma } \, , \nonumber \\
r_{20} &=& \frac{\delta \, r_{11}+ \epsilon \, r_{12} + \sigma \, r_{13} }{\gamma },\, r_{30} = \frac{\beta \, r_{10}}{\gamma }, \, r_{21} = \frac{\delta \, r_{10}}{\gamma }, \, r_{22} = \frac{\epsilon \, r_{10} }{\gamma }, \, r_{23} = \frac{\sigma \, r_{10} }{\gamma } \, . 
\end{eqnarray}
Notice that only $8$ parameters were determined and thus one has $7$ free variables $ r_{10}, r_{01}$, $r_{02}, r_{03}$, $r_{11}, r_{12}$, $r_{13} $. Hence, from Table~\ref{tabdim} and Eq.~(\ref{dimn}), it implies that $ \hat{H} $ is double degenerate. 

\subsubsection{Pure state solution}
\hfill

For the pure state case, one has to solve the system of equations $ \hat{\rho}^2=\hat{\rho} $. Finally one has $2$ free parameters $r_{02}$ and $r_{03}$. One possibility is to solve $r_{02}=r_{03}=0$ and so one gets four solutions for the extremal density matrices denoted by $ \{ \hat{\rho}^{c}_{1 \pm} ,\, \hat{\rho}^{c}_{2 \pm} \} $, whose respective Bloch vectors, Correlation and Schlienz-Mahler matrices are 
\begin{eqnarray}
\tau^A_{1 \pm} & = & \pm \biggl(  \frac{\gamma }{E} , \, -\frac{\delta }{E} \frac{\Delta }{\sqrt{\Delta ^2 + \omega ^2}} , \,  \frac{\beta }{E} \biggr) , \quad
\tau^B_{1 \pm} = \biggl(  -\frac{\Delta }{\sqrt{\Delta ^2 + \omega ^2}} , \,0 , \,  0 \biggr) \, , \nonumber \\
\boldsymbol{C}_{1 \pm} & = & \pm \frac{1}{E \, \Delta \, \sqrt{\Delta ^2+\omega ^2 }} \left(
\begin{array}{cccc}
- \gamma (\Delta ^2 + \omega ^2) &  \gamma  \delta  \epsilon \pm \beta  E \sigma &   \gamma  \delta  \sigma \mp \beta  E \epsilon   \\
 \delta \, \Delta \sqrt{\Delta ^2+\omega ^2 } &  \epsilon \, \Delta \sqrt{\Delta ^2+\omega ^2 } & \sigma \, \Delta \sqrt{\Delta ^2+\omega ^2 } \\
- \beta (\Delta ^2 + \omega ^2) & \beta  \delta  \epsilon \mp \gamma  E \sigma  & \beta  \delta  \sigma \pm \gamma  E \epsilon 
\end{array}
\right) \, , \nonumber \\
\boldsymbol{M}_{1 \pm} & = & \boldsymbol{C}_{1 \pm} \pm \frac{1}{E \, \Delta \, \sqrt{\Delta ^2+\omega ^2 }} \left(
\begin{array}{ccc}
- \gamma \, \Delta^2 & 0 & 0 \\
\frac{\delta \, \Delta^3}{\sqrt{\Delta ^2+\omega ^2 }} & 0 & 0 \\
- \beta \, \Delta^2 & 0 & 0 \\
\end{array}
\right)  \, , \label{solsdeg} \\
\tau^A_{2 \pm} & = &  \pm \biggl(  \frac{\gamma }{E} , \, \frac{\delta }{E} \frac{\Delta }{\sqrt{\Delta ^2 + \omega ^2}} , \,  \frac{\beta }{E} \biggr) , \quad
\tau^B_{2 \pm} = \biggl(  \frac{\Delta }{\sqrt{\Delta ^2 + \omega ^2}} , \,0 , \,  0 \biggr) \, , \nonumber  \\
\boldsymbol{C}_{2 \pm} & = & \pm \frac{1}{E \, \Delta \, \sqrt{\Delta ^2+\omega ^2 }} \left(
\begin{array}{cccc}
 \gamma (\Delta ^2 + \omega ^2) & -( \gamma  \delta  \epsilon \pm \beta  E \sigma ) &  -( \gamma  \delta  \sigma \mp \beta  E \epsilon )  \\
 \delta \, \Delta \sqrt{\Delta ^2+\omega ^2 } &  \epsilon \, \Delta \sqrt{\Delta ^2+\omega ^2 } & \sigma \, \Delta \sqrt{\Delta ^2+\omega ^2 } \\
 \beta (\Delta ^2 + \omega ^2) & -( \beta  \delta  \epsilon \mp \gamma  E \sigma ) & - ( \beta  \delta  \sigma \pm \gamma  E \epsilon )
\end{array}
\right) \, , \nonumber \\
\boldsymbol{M}_{2 \pm} & = & \boldsymbol{C}_{2 \pm} \pm \frac{1}{E \, \Delta \, \sqrt{\Delta ^2+\omega ^2 }} \left(
\begin{array}{ccc}
 \gamma \, \Delta^2 & 0 & 0 \\
 \frac{ \delta \, \Delta^3}{\sqrt{\Delta ^2+\omega ^2}} & 0 & 0 \\
  \beta \, \Delta^2  & 0 & 0 \\
\end{array}
\right)  \, , \nonumber
\end{eqnarray}
where we define $ \Delta = \sqrt{\beta ^2+\gamma ^2} $, $\omega = \sqrt{ \sigma ^2+\epsilon ^2 } $, and $ E = ( \det \hat{H} )^{\frac{1}{4}} = \sqrt{ \Delta^2 + \omega^2 + \delta^2 } $. 

The respective expectation values of the Hamiltonian are
\begin{eqnarray}
{\rm Tr} ( \hat{\rho}^{c}_{1 \pm} \,\hat{H} ) = \pm E \, , \quad
{\rm Tr} ( \hat{\rho}^{c}_{2 \pm} \,\hat{H} ) = \pm E \, , \nonumber
\end{eqnarray}
and it is possible to corroborate that the set $ \{ \hat{\rho}^c_{1\pm}, \, \hat{\rho}^c_{2\pm} \} $ constitutes a complete set of orthogonal rank one projectors because $ \hat{\rho}^c_{1+} + \hat{\rho}^c_{1-} + \hat{\rho}^c_{2+} + \hat{\rho}^c_{2-} = \hat{I}_{4}$. 

From~(\ref{betapar}), the parameter $ \beta $ for each state of $ \{ \hat{\rho}^c_{1\pm}, \, \hat{\rho}^c_{2\pm} \} $ take the same value,
\begin{eqnarray}
\beta & = & \frac{16}{15} \, S_L ( 1 + S_L )  ,  \quad
S_L = \frac{\omega^2}{ 2( \Delta^2 + \omega^2 ) } \, , \label{betaec}
\end{eqnarray}
indicating entanglement between the qubits for a collection of finite values of  $ \{ \beta , \, \gamma , \, \sigma , \, \epsilon \} $. The defined $ S_L $ is precisely the Linear Entropy, 
\begin{eqnarray}
S_L = \frac{1}{2} \biggl( 1 - |\, \tau^A_{1 \pm} \, |^2 \biggr) = \frac{1}{2} \biggl( 1 - | \, \tau^B_{1 \pm} \, |^2 \biggr) \, .
\end{eqnarray}
One can observe that this quantity is independent of the parameter $ \delta$ and there is no entanglement if $ \sigma = \epsilon = 0$. 

\subsubsection{Mixed state solution}
\hfill

The general extremal mixed state of the Hamiltonian can be written in the form
\begin{eqnarray}
\hat{\rho}_{mix} = \mathcal{P}_1 \, \hat{\rho}^{c}_{1 +} + \mathcal{P}_2 \, \hat{\rho}^{c}_{2 +} + \mathcal{P}_3 \, \hat{\rho}^{c}_{1 -} + \mathcal{P}_4 \, \hat{\rho}^{c}_{2 -} \, , \label{mixden}
\end{eqnarray}
where $ \sum_j^4 \mathcal{P}_j = 1 $ and $ 0 \leq \mathcal{P}_j \leq 1 $. By means of the Fano representation~(\ref{fanoform}) and the solutions~(\ref{solsdeg}), it is straightforward that
\begin{eqnarray}
\hat{\rho}_{mix} &=& \frac{1}{4} \left( \hat{I}_4 + \sum_{p=1}^{3} \tau^A_p \, \hat{D}_{p,0} +  \sum_{q=1}^{3} \tau^B_q \,  \hat{D}_{0,q} + \sum_{p=1}^{3} \sum_{q=1}^{3} C_{p\,q} \,  \hat{D}_{p,q}  \right) \, , \label{mixdenfano}
\end{eqnarray}
where
\begin{eqnarray}
\tau^A & = &  \biggl(  \frac{ x \, \gamma }{E} , \, -\frac{\delta }{E} \frac{z\, \Delta }{\sqrt{\Delta ^2 + \omega ^2}} , \,  \frac{x \, \beta }{E} \biggr) , \quad
\tau^B = \biggl(  -\frac{y \, \Delta }{\sqrt{\Delta ^2 + \omega ^2}} , \,0 , \,  0 \biggr) \, , \nonumber \\
\boldsymbol{C} & = & \frac{1}{E \, \Delta \, \sqrt{\Delta ^2+\omega ^2 }} \left(
\begin{array}{cccc}
- z\, \gamma (\Delta ^2 + \omega ^2) & z\, \gamma  \delta  \epsilon + y\, \beta  E \sigma &  z\, \gamma  \delta  \sigma - y\, \beta  E \epsilon   \\
x\, \delta \, \Delta \sqrt{\Delta ^2+\omega ^2 } & x\, \epsilon \, \Delta \sqrt{\Delta ^2+\omega ^2 } & x\, \sigma \, \Delta \sqrt{\Delta ^2+\omega ^2 } \\
- z\, \beta (\Delta ^2 + \omega ^2) & z\, \beta  \delta  \epsilon - y\, \gamma  E \sigma  & z\, \beta  \delta  \sigma + y\, \gamma  E \epsilon 
\end{array}
\right) \, , \label{solsdegmix} \\
\boldsymbol{M} & = & \boldsymbol{C} - \frac{1}{E \, \Delta \, \sqrt{\Delta ^2+\omega ^2}} \left(
\begin{array}{ccc}
\frac{ x\, y \, \gamma \, \Delta^2  }{\sqrt{\Delta ^2+\omega ^2}} & 0 & 0 \\
\frac{- y\, z \, \delta \, \Delta^3}{\sqrt{\Delta ^2+\omega ^2}} & 0 & 0 \\
 x\, y \, \beta \, \Delta^2   & 0 & 0 \\
\end{array}
\right)  \, , \nonumber
\end{eqnarray}
with the following definitions $ x = \mathcal{P}_1 + \mathcal{P}_2 - \mathcal{P}_3 - \mathcal{P}_4 $, $ y = \mathcal{P}_1 - \mathcal{P}_2 + \mathcal{P}_3 - \mathcal{P}_4 $ and $ z = \mathcal{P}_1 - \mathcal{P}_2 - \mathcal{P}_3 + \mathcal{P}_4 $. The respective determinants of $ \boldsymbol{C} $ and $ \boldsymbol{M} $ are
\begin{eqnarray}
\det \boldsymbol{C} & = &  - \, \frac{ \omega^2 }{\Delta ^2+\omega ^2} \; x \, y \, z \, , \label{detcmix} \\
\det \boldsymbol{M} & = &  \frac{\omega ^2}{ E^2 \left(\Delta ^2+\omega ^2\right)^2} \, \biggl( \Delta ^2 (\Delta ^2+\omega ^2 ) x^2 y^2 + \delta ^2 \Delta ^2 y^2 z^2 - E^2 (\Delta ^2+\omega ^2) x y z \biggr) \, . \label{detmmix}
\end{eqnarray}

Now we are going to use the PPT criterion by considering the expression~(\ref{pospt}). Thus in Table~\ref{tab:1}, for any extremal mixed state~(\ref{mixdenfano}), the set of values $(a^{PT}_2, a^{PT}_3, a^{PT}_4)$ from Eqs.~(\ref{pospt}) are shown. Therefore we have five possibilities for the extremal mixed density matrices. Three of the cases determine entangled extremal density matrices, which are given for  three equal probabilities, i.e.,  $\mathcal{P}_2=\mathcal{P}_3=\mathcal{P}_4$, two equal probabilities $\mathcal{P}_3=\mathcal{P}_4$, and all probabilities different. The separable cases occur for the maximal mixed state and when one has two equal probabilities.  For all the cases we have similar results permutating the values of the probabilities.

The PPT criterion can be also applied to the pure case taking $\mathcal{P}_2=\mathcal{P}_3=\mathcal{P}_4=0$. For this we have $\det \boldsymbol{C} = -\omega^2/(\Delta^2+\omega^2)$ and $\det \boldsymbol{M}= \omega^2 \, \det \boldsymbol{C}$, in agreement with the discussion of the previous subsection.
\begin{table*}[ht!]
\centering
\caption{ Values $ \{ a^{PT}_2 ,\, a^{PT}_3 ,\, a^{PT}_4 \} $ from Eqs.~(\ref{pospt}) in terms of the probability coefficients $ \{ \mathcal{P}_1 , \mathcal{P}_2 , \mathcal{P}_3 , \mathcal{P}_4 \} $ of the general 2-qubit mixed extremal density matrix~(\ref{mixden}). Five cases are taken into account according to the different strata shown in the Table~\ref{tabdim} for $d=4$. } \label{tab:1}
\medskip
\resizebox{0.98  \textwidth}{!} {
\begin{tabular}{ | c || c | c | c |}
\hline
&&&\\[-2mm]
$ \boldsymbol{(\mathcal{P}_1 , \mathcal{P}_2 , \mathcal{P}_3 , \mathcal{P}_4)} $ &  $\boldsymbol{ a^{PT}_2 } $  & $ \boldsymbol{ a^{PT}_3 } $ & $ \boldsymbol{ a^{PT}_4 } $  \\[2mm]
\hline 
\hline
&&&\\[-2mm]
$ ( 1/4 , 1/4 , 1/4, 1/4 ) $  & $ 3/8 $ & $ 1/16 $ &  $ 1/256 $   \\[3mm]
\hline
&&&\\[-2mm]
$ ( 1 - 3\, b , b , b, b ) $ & $ 3(1-2 b) b $ & $ (3 - 8 b) b^2 - \, \frac{ \omega ^2 }{4 (\Delta ^2+\omega ^2)}(1-4 b)^3  $ & $ (1 - 3 b) b^3 $ + $ \frac{\omega ^2}{16 \, E^2 \left(\Delta ^2+\omega ^2\right)^2} \biggl( (1-4 b)^3 \biggl( (1-4 b) \delta ^2 \Delta ^2 - $  \\
 & &  &  $ ( \Delta ^2+\omega ^2 ) ( (4 b-1) \Delta ^2+E^2 ) \biggr) \biggr) $ \\[3mm]
\hline
&&&\\[-2mm]
$ ( 1/2 - b , 1/2 - b , b, b ) $ & $ \frac{1}{4}+ b -2 b^2 $ & $ \frac{1}{2} (1-2 b) b $ &  $ \left(b-\frac{1}{2}\right)^2 b^2 $   \\[3mm]
\hline
&&&\\[-2mm]
$ ( 1-b-2\, c , b , c, c ) $ & $ -b^2-2 b c+ $ & $ c \biggl(c-4 b c -2(b-1)b-2c^2 \biggr) $ &  $ b c^2 (1-b - 2 c) $ + $ \frac{\omega ^2}{16 \, E^2 \left(\Delta ^2+\omega ^2\right)^2} \biggl( (2 b+2 c-1)^2 \biggl(  \delta ^2 \Delta ^2 (2 b+2 c-1)^2 + $ \\
& $ b+c (2-3 c) $ & $ - \, \frac{ \omega ^2 }{4 (\Delta ^2+\omega ^2)}  (1-4 c) (2 b+2 c-1)^2 $ &  $ (4 c-1) (\Delta ^2+\omega ^2 ) ((4 c-1) \Delta ^2+E^2 ) \biggr) \biggr) $ \\[3mm]
\hline
&&&\\[-2mm]
$ ( 1-b-c-d ,b , c, d ) $ & $ -b^2-d (b+c)-b c+ $ &  $ b (1-c-d) (c+d)+ c d (1-c-d) -b^2(c+d) $ &  $ b c d (1-b-c-d) $ + $ \frac{\omega ^2}{16 \, E^2 \left(\Delta ^2+\omega ^2\right)^2} \biggl( \Delta ^2 (\Delta ^2+\omega ^2 ) (1 - 2 c - 2 d)^2 (1 - 2 b - 2 d)^2 +  $ \\
  & $ b-c^2+c-d^2+d $ & $ - \, \frac{ \omega ^2 }{4 (\Delta ^2+\omega ^2)} (1 - 2 b - 2 c ) (1 - 2 b - 2 d ) (1 - 2 c - 2 d ) $ &   $ \delta ^2 \Delta ^2 (1 - 2 b - 2 d)^2 (1 - 2 b - 2 c)^2 - $ \\
  & &  &  $  E^2 (\Delta ^2+\omega ^2) (1 - 2 c - 2 d) (1 - 2 b - 2 d) (1 - 2 b - 2 c) \biggr) $ \\
\hline
\end{tabular}}
\end{table*}

\subsubsection{States with Kramers invariance}
\hfill

From the 2-qubit degenerate case Hamiltonian studied above, we have found that pure extremal states do not commute with the time-reversal operator $ \hat{T} $. Due to this observation, we are going to prove that, in general, this occurs for any $2 N$ dimensional Hilbert space.

{\bf Proposition 1.} If $ [ \hat{T} ,\hat{H} ] = \boldsymbol{0} $, extremal pure states do not commute with $\hat{T}$.

{\it Proof.} Consider that $ [ \hat{T} , \hat{H} ] = \boldsymbol{0} $. Consequently, for $ k=1,\ldots,N $, one can construct an orthonormal basis consisting on eigenvectors of $ \hat{H} $ as $\{ \, | \psi_1 \rangle , | \hat{T} \psi_1 \rangle$, $| \psi_2 \rangle, | \hat{T} \psi_2 \rangle$, $\ldots, | \psi_N \rangle$, $| \hat{T} \psi_N \rangle \, \}$,
such that the pure states 
$$ \hat{\rho}_k = \hat{\rho}_k^2 = | \psi_k \rangle \langle \psi_k |  \, ,\quad \hat{\eta}_k = \hat{\eta}_k^2 = \hat{T} |  \psi_k \rangle \langle \psi_k  | \hat{T}^{-1}  $$
commuting with $\hat{H}$ are rank-one orthogonal projectors (Kramers pairs) which describe a two-dimensional degenerate space of $\hat{H}$. Suppose now that $ [ \hat{T} ,  \hat{\rho}_k ] = \boldsymbol{0} $. Then, it follows that $ {\rm Tr} ( \hat{\rho}_k \, \hat{\eta}_k ) = {\rm Tr} ( \hat{\rho}_k \, \hat{\rho}_k ) = 1 $, contrary to the orthogonality of the basis. Because all pure states are unique up to a unitary transformation, the conclusion holds. {\it q.e.d.}

On the other hand, rank-two projectors constructed by Kramers pairs commute with both $ \hat{H} $ and $ \hat{T} $.

{\bf Proposition 2.} If $ [ \hat{T} ,\hat{H} ] = \boldsymbol{0} $, rank-two projectors formed by Kramers pairs commute with $\hat{T}$.

{\it Proof.} By means of the Kramers pairs $ \{ \hat{\rho}_k \, ,\, \hat{\eta}_k \} $ defined above, one can construct rank-two projectors as $ \hat{P}_{k}=\hat{\rho}_k+\hat{\eta}_k$, which satisfies $ \hat{P}_{k}^2=\hat{P}_{k} $. Then, 
\begin{eqnarray}
[ \hat{T} ,\hat{P}_{k} ] & = & \hat{T}\,\hat{\rho}_k - \hat{\rho}_k \, \hat{T} + \hat{T}\,  (\hat{T}\, \hat{\rho}_k \, \hat{T}^{-1} ) - (\hat{T}\, \hat{\rho}_k \, \hat{T}^{-1} ) \, \hat{T} \, , \nonumber \\
 & = & - \hat{\rho}_k ( \hat{T} + \hat{T}^{-1} ) = \boldsymbol{0} \, , \nonumber
\end{eqnarray}
where it was used that $\hat{T}^2=-\hat{I}$ plus $ \hat{T}\,\hat{T}^{-1} =\hat{I}$ imples $  \hat{T} + \hat{T}^{-1} = \boldsymbol{0} $. {\it q.e.d.}

Consequently, in addition of being separable states, the maximal mixed state, mixed states with $(\mathcal{P}_1 , \mathcal{P}_2 , \mathcal{P}_3 , \mathcal{P}_4) = ( 1/2 - b , 1/2 - b , b, b) $ and its permutations, possess Kramers invariance.

\section{Summary and Conclusions}\label{con}


We provide a self-contained method to determine the extremal density matrices of a finite dimensional time-reversal Hamiltonian. These extremal states commute with the Hamiltonian operator and optimise its mean value, such that the conventional variational principle is extended to mixed states. 

We also establish a novel procedure to analyze the entanglement of extremal density matrices. It has the advantage of reaching any desirable extremal state, either separable or entangled, by changing the parameters of the Hamiltonian. It was applied for two families of cases of the 2-qubit Hamiltonian, in which, by varying its parameters, it is possible to keep it non degenerate or to acquire Kramers degeneracy. For the non degenerate case, we show that their associated extremal pure and mixed states are separable. When the Hamiltonian exhibits Kramers degeneracy, we have found both possibilities for extremal pure states, depending on the parameters of the Hamiltonian. For the extremal mixed matrices we also have both possibilities classified in five cases according to their eigenvalues degeneracy. 

The sufficiency of the PPT criterion in the qubit-qubit and qubit-qutrit systems makes that our procedure has no ambiguities and it can be applied to any observable by replacing the Hamiltonian. In higher dimensions, it can be implemented with the respective consideration that the PPT criterion is just a necessary condition. We want to enhance that our procedure encompasses Hamiltonian and states in the same context, thus, it is possible to discuss how the Hamiltonian degeneracy, its symmetry, the purity and entanglement of states in finite dimensional Hilbert space are intertwined. This is neither clear nor direct in the context of the diagonalisation procedure using the secular equation.

In comparison with other separability criteria, the advantage of linking the extremal density matrices method with the PPT criterion is the algebraic aspect of the approach, i.e., the posivity conditions on the partial transposed density matrix (inequalities (\ref{pospt})) separate in explicit way the regions for which there will be or not entanglement, for both pure and mixed states, by varying the parameters of the Hamiltonian. Consequently, this makes the procedure general and simple, without the need to introduce extra concepts. 

\appendix

\section{Positivity conditions for the Density Operator} \label{B} 

The characteristic polynomial $P_d(x)$ for the density matrix acting on a $d$-dimensional Hilbert space is given by
\begin{equation}\label{ec:eq50}
P_d(x) \equiv det(x \hat{I}_d - \hat{\rho}) = \sum^d_{j=0} (-1)^j a_j x^{d-j} = 0 \, , 
\end{equation}
with the definitions $a_0 = a_1 \equiv 1$ and $a_d = \det \hat{\rho}$. For $d \geq 2$, the real coefficients $ \{ a_k \} $ are bounded by~\cite{deen,byrd}
\begin{equation}\label{s2:eq1}
0 \leq a_k \leq \frac{1}{d^k} {d \choose k} \, ,
\end{equation}  
where ${d \choose k}$ denotes a binomial coefficient. The upper bound defines the most mixed state with maximum von Neumann entropy, while the lower bound specifies pure states which has zero entropy. Additionally, it is known that $ a_k =0 $ for all values of $k > {\rm rank }( \hat{\rho})$~\cite{tapia}. 

The coefficients $ \{ a_k \} $ can be obtained by means of the Girard-Waring formula~\cite{tapia,gould}
\begin{eqnarray}
a_k = \sum^{p(k)}_{i=1} \, \prod^{k}_{j=1} \frac{(-1)^{(j-1) q_{i j}}}{j^{q_{i j}} \: q_{i j}\, ! } \, \left( t_j \right)^{q_{i j}} \, , \label{giwar}
\end{eqnarray}
where $ t_j \equiv {\rm Tr} (\hat{\rho}^j )$, for $j=1,\ldots , d $ and $ \{ q_{i j} ,\, p(k) \} $ denote, respectively, the natural numbers solutions and (the partition function $p(k)$) the number of solutions, without regarding to order, of the linear Diophantine equation 
\begin{eqnarray}
1\, q_{i 1} + 2 \, q_{i 2} + 3 \, q_{i 3} + \cdots + k \, q_{i k} = k \, , \label{dioph}
\end{eqnarray}
with $ i = 1, 2, \ldots, p(k) $. For the first four values of $ k $, $ p(k) $ and $ \{ q_{i j} \} $ are given by
\begin{eqnarray}
p(k=1)=1  & \Rightarrow &   q_{1 1} = 1  \, , \nonumber \\
p(k=2)=2   & \Rightarrow & \bigl\{ \, ( q_{1 1} ,  q_{1 2} ) = (2, 0) , \quad ( q_{2 1} ,  q_{2 2} ) = (0, 1) \, \bigr\}  \, , \nonumber  \\
p(k=3)=3  & \Rightarrow &
\Biggl\{
\begin{array}{cc}
( q_{1 1} ,  q_{1 2} ,  q_{1 3} ) = (3, 0, 0), & ( q_{2 1} , q_{2 2} , q_{2 3} ) = (1, 1, 0), \\
( q_{3 1} , q_{3 2} ,  q_{3 3} ) = (0, 0, 1)  &
\end{array}
\Biggr\} \, , \label{diophsol}  \\
p(k=4)=5  & \Rightarrow & 
\left\{
\begin{array}{cc}
( q_{1 1} ,  q_{1 2} ,  q_{1 3} , q_{1 4} ) = (4, 0, 0, 0), & ( q_{2 1} ,  q_{2 2} ,  q_{2 3} ,  q_{2 4} ) = (2, 1, 0, 0) , \\
( q_{3 1} , q_{3 2} ,  q_{3 3} ,  q_{3 4} ) = (0, 2, 0, 0), & ( q_{4 1} , q_{4 2} ,  q_{4 3} ,  q_{4 4} ) = (1, 0, 1, 0) , \\
( q_{5 1} ,  q_{5 2} ,  q_{5 3} , q_{5 4} ) = (0, 0, 0, 1) 
\end{array}
\right\} \, . \nonumber 
\end{eqnarray}

The inverse of the Girard-Waring formula (\ref{giwar}) exists and its given by~\cite{gould}
\begin{eqnarray}
t_k = k \; \sum^{p(k)}_{i=1} \, \biggl(  M_i -1 \biggr) \, ! \, \prod^{k}_{j=1}  \, \frac{ (-1)^{(j-1) q_{i j}}}{ q_{i j}\, ! } \, \left( a_j \right)^{q_{i j}}   \, , \label{giwarinv}
\end{eqnarray} 
where $ M_i \equiv \sum_{s=1}^k \, q_{is} $. 

The formulas (\ref{giwar}) and (\ref{giwarinv}) can also be expressed in Plemelj-Smithies form~\cite{gohberg},
\begin{eqnarray}
a_k = \frac{1}{k \, !} \left| \begin{array}{ccccc}
	t_1 &	1 &	0 &  \ldots & 0 \\
	t_2 &	t_1 &	2 &  \ldots & 0 \\
	\vdots &	\vdots &	\ddots & \ddots & \vdots \\
	t_{k-1} &	t_{k-2} &	t_{k-3} & \ddots & k-1 \\
	t_{k} &	t_{k-1} &	t_{k-2} & \ldots & t_1
 \end{array} \right| , \quad
t_k = \left| \begin{array}{ccccc}
	a_1 &	1 &	0 &  \ldots & 0 \\
	2 a_2 &	a_1 &	1 &  \ldots & 0 \\
	\vdots &	\vdots & \ddots & \ddots & \vdots \\
	(k-1) a_{k-1} &	a_{k-2} &	a_{k-3} &  \ddots & 1 \\
	k a_{k} &	a_{k-1} &	a_{k-2} & \ldots & a_1
 \end{array} \right| \, . \nonumber
\end{eqnarray}

The density matrix must satisfy the following three properties: (a) It  has trace one; (b) all its eigenvalues are positive or zero; and (c) it is Hermitian. Given a monic real polynomial, the inverse problem of deciding when it comes from a density matrix requires these assumptions being translated into polynomial conditions. In other words, the trace-one requisite is equivalent to $a_0 = a_1 = 1$, the semi-positivity condition implies that expressions~(\ref{s2:eq1}) must be fulfilled, and the hermiticity condition is taken into account through the Bezoutian matrix $ \boldsymbol{B}_d $, i.e., a polynomial with real coefficients has reals roots iff $ \boldsymbol{B}_d $ is positive definite~\cite{procesi}. In terms of $ t_j \equiv {\rm Tr} (\hat{\rho}^j ) $, with $j=1,\ldots , d $, the symmetric Bezoutian matrix is defined by~\cite{procesi,schwarz,gerdt}
\begin{equation} 
\boldsymbol{B}_d = \left(
\begin{array}{ccccc}
	d      & t_1  	& t_2 		& \cdots 	& t_{d-1} 	\\
	t_1    & t_2  	& t_3 		&  \ddots 	& t_d   	\\
	t_2    & t_3   	& \ddots   	&        	& t_{d+1}   \\
	\vdots & \ddots &     		&        	& \vdots     \\
	t_{d-1} & t_d 	& t_{d+1} 	& \cdots 	& t_{2(d-1)}
\end{array} \right) \, . \label{B:eq1}
\end{equation}

For a given monic real polynomial in $x$ of degree $d$, $P_d(x)$, having roots $ \{ \mathcal{P}_1 , \mathcal{P}_2 , \ldots, \mathcal{P}_d \} $, its associated Bezoutian matrix $\boldsymbol{B}_d $ has the following properties:
\begin{enumerate}[a)]
\item The rank of the Bezoutian equals the number of distinct roots of $P_d(x)$~\cite{procesi}.
\item (Sylvester criteria). The number of real roots of $P_d(x)$ equals the signature (the difference between positive and negative real roots) of its Bezoutian~\cite{procesi}.
\item (Reality condition). $P_d(x)$ has all its roots real and distinct iff the Bezoutian matrix is positive definite~\cite{procesi}.
\item In the case of $d=2,3 $, $ \det \boldsymbol{B}_d $ is the only positivity condition of the Bezoutian~\cite{niculescu}.
\item (Degeneracy condition). The discriminant of $P(x)$ is equal to the determinant of the Bezoutian, thus the condition for repeated roots of $P(x)$ is obtained by the vanishing of $ \det \boldsymbol{B}_d $~\cite{bhattacharya}.  
\item By means of the Vandermonde matrix,
\begin{eqnarray}
\boldsymbol{V}_d &=& \left(
\begin{array}{cccc}
 1 & 1 & \cdots & 1 \\
 \mathcal{P}_1 & \mathcal{P}_2 & \cdots & \mathcal{P}_d \\
 \mathcal{P}^2_1 & \mathcal{P}^2_2 & \cdots & \mathcal{P}^2_d \\
 \vdots & \vdots & \ddots & \vdots \\
 \mathcal{P}^{d-1}_1 & \mathcal{P}^{d-1}_2 & \cdots & \mathcal{P}^{d-1}_d \\
\end{array}
\right) \, ,
\end{eqnarray}
the Bezoutian is obtained as $\boldsymbol{B}_d = \boldsymbol{V}_d \, \boldsymbol{V}^T_d$.
\end{enumerate}

Combining all above results, a monic real polynomial coming from a density matrix must satisfy the following system of $d-1$ simultaneous polynomial equations:
\begin{equation}\label{s2:eq3}
c_k = a_k \, , \qquad {\rm for} \quad k=2,3,\ldots,d \, ,
\end{equation}
where constants $c_k$ fix the degree of mixing of the system, and the compatible region among them is obtained with the intersection of the semi-positivity conditions of the density matrix from~(\ref{s2:eq1}) with the respective positivity conditions of the Bezoutian matrix~\cite{gerdt}. 
%

For instance, taking into account the equality~(\ref{s2:eq3}), the semi-positivity conditions~(\ref{s2:eq1}) of the density matrices with dimensions $d=4$ are given by
\begin{eqnarray}
&& 0 \leq c_2  \leq \frac{3}{8} \, , \nonumber \\
&& 0 \leq c_3  \leq \frac{1}{16} \, , \label{B:eq7} \\
&& 0 \leq c_4  \leq \frac{1}{256} \, , \nonumber
\end{eqnarray}
and the respective Bezoutian matrix is
\begin{equation*}
\boldsymbol{B}_4 =  \left(
\begin{array}{cccc}
	4		& 1		& t_2 	& t_3	\\
	1    	& t_2  	& t_3	& t_4	\\
	t_2		& t_3 	& t_4	& t_5    \\
	 t_3	& t_4	& t_5	& t_6
\end{array} \right) \, ,
\end{equation*} 
where, the relations between $ \{ a_p \} $ with $ \{ t_k \} $ given in Eq.~(\ref{giwar}) yields,
\begin{eqnarray}
t_2 & = & 1 - 2 \, c_2 \, , \nonumber \\
t_3  & = & 1 - 3 c_2 + 3c_3 \, , \nonumber \\
t_4  & = &  2 (c_2-2) c_2 +4 c_3 - 4 c_4 +1 \, , \label{B:eq8} \\
t_5 & = & 5 c_2 (c_2 -c_3 -1)+5 c_3 +5 c_4 +1 \, , \nonumber \\
t_6 & = & 9 c_2^2 - 2 c_2^3 -6 (2 c_3 + c_4 +1) c_2 +3 c_3 (c_3 +2)+6 c_4 +1 \, . \nonumber
\end{eqnarray}

All the positivity conditions on $ \boldsymbol{B}_4  $ are
\begin{eqnarray}
{\rm Tr} \boldsymbol{B}_4  & \geq & 0 \, , \nonumber \\
  \frac{1}{2} \biggl( ({\rm Tr} \boldsymbol{B}_4)^2 - {\rm Tr} \boldsymbol{B}_4^2 \biggr) & \geq & 0  \, , \label{B:eq9a}  \\ 
 \frac{1}{6} \biggl( ({\rm Tr} \boldsymbol{B}_4)^3 - 3 {\rm Tr} \boldsymbol{B}_4 \, {\rm Tr} \boldsymbol{B}_4^2 + 2 {\rm Tr} \boldsymbol{B}_4^3 \biggr) & \geq & 0 \, , \nonumber \\
  \det \boldsymbol{B}_4 & \geq & 0  \, , \nonumber
\end{eqnarray}
where the last one is the main condition. Nevertheless, the remaining ones are crucial to avoid fake points in the compatible region for $ \{ c_2,\, c_3,\, c_4 \}$. 

Hence, for the set $ \{ c_2,\, c_3,\, c_4 \}$, the region which satisfies the inequalities system formed by~(\ref{B:eq7}) and~(\ref{B:eq9a}), is shown in Fig.~\ref{region}.  Notice that, by making zero $c_4$, we obtain the $d=3$ result, while by making zero two eigenvalues of the density matrix the line associated to the case $d=2$ is obtained ($c_3=c_4=0$). Inside the solid figure (orange color) one has the solution for $4$ eigenvalues of the density matrix different from zero, whereas the surfaces are associated to $2$ degenerated eigenvalues (blue color). The curve for the case with three equal eigenvalues and the other different is also shown (green color).

{\bf Acknowledgments.} This work was partially supported by CONACyT-M\'exico (under Project No.~238494). 


\end{document}